\documentclass[aps, prb,twocolumn,showpacs,preprintnumbers,amsmath,amssymb]{revtex4}
\usepackage{mathrsfs}

\usepackage{amsmath}
\usepackage{amssymb}
\usepackage{graphicx}
\usepackage{color}
\usepackage{dcolumn}
\usepackage{bm}
\makeatletter

\newcommand{\Rmnum}[1]{\expandafter\@slowromancap\romannumeral #1@}
\makeatother

\begin{document}

\title{Magnetic property and crystalline electric field effect in ThCr$_2$Si$_2$-type CeNi$_2$As$_2$}

\author{Yongkang Luo$^{1}$, Jinke Bao$^{1}$, Chenyi Shen$^{1}$, Jieke Han$^{1}$, Xiaojun Yang$^{1}$, Chen Lv$^{1}$, Yuke Li$^{2}$, Wenhe Jiao$^{1}$, Bingqi Si$^1$, Chunmu Feng$^{1}$, Jianhui Dai$^{2}$, Guanghan Cao$^{1}$, and Zhu-an Xu$^{1}$\footnote[1]{Electronic address: zhuan@zju.edu.cn}}

\address{$^1$Department of Physics and State Key Laboratory of Silicon Materials, Zhejiang University,
Hangzhou 310027, China,}
\address{$^2$Department of Physics, Hangzhou Normal University,
Hangzhou 310036, China.}

\date{\today}

\begin{abstract}

Millimeter sized ThCr$_2$Si$_2$-type CeNi$_2$As$_2$ single crystal
was synthesized by NaAs flux method and its physical properties
were investigated by magnetization, transport and specific heat
measurements. In contrast to the previously reported
CaBe$_2$Ge$_2$-type CeNi$_2$As$_2$, the ThCr$_2$Si$_2$-type
CeNi$_2$As$_2$ is a highly anisotropic uniaxial antiferromagnet
with the transition temperature $T_N$=4.8 K. A magnetic field
induced spin flop transition was seen below $T_N$ when the applied
$\textbf{B}$ is parallel to the $\textbf{c}$-axis, the magnetic
easy axis, together with a huge frustration parameter
$f=\theta_W/T_N$. A pronounced Schottky-like anomaly in specific
heat was also found around 160 K, which could be attributed to the
crystalline electric field effect with the excitation energies
being fitted to $\Delta_1=$325 K and $\Delta_2=$520 K,
respectively. Moreover, the in-plane resistivity anisotropy and
low temperature X-ray diffractions suggest that this compound is a
rare example exhibiting a possible structure distortion induced by
the $4f$-electron magnetic frustration.
\end{abstract}


\maketitle

\section{Introduction}

The interest in the ThCr$_2$Si$_2$-type structure has been
rekindled since the discovery of superconductivity (SC) in
(Ba$_{1-x}$K$_x$)Fe$_2$As$_2$\cite{Rotter}. SC was also achieved
when Ba is replaced by other alkaline earths like Ca and Sr or
even the divalent rare earth Eu, either by chemical doping or
pressure
effect\cite{Park-Ca122,Patricia-Sr/Ba122,Sefat-Ba122_Co,XuZA-Ba122_Ni,CaoGH-Eu122_P,ChenXH-Eu122_Co,CaoGH-Eu122_Ru}.
On the other hand, the nickel based pnictide, e.g. BaNi$_2$As$_2$,
was reported to show SC too, although the nickels are non-magnetic
and the $T_c$ is much lower\cite{Sefat-BaNi2As2}. In these
122-compounds formulated with $ATm_2$As$_2$ ($A$=Ca, Sr, Ba or Eu,
$Tm$=transition metals), two vertically reversed $Tm$As layers are
sandwiched along $c$-axis, while the $A$ atoms are embedded in
between, following a sequence of $Tm$As-$A$-$Tm$As. This
crystalline structure constitutes a platform for understanding the
interplay between Kondo interaction and
Ruderman-Kittel-Kasuya-Yosida (RKKY) interaction if $A$ is
replaced by magnetic rare earths. Indeed, the research for the
$4f$-electron correlation in ThCr$_{2}$Si$_2$ structured compounds
has been a long story, and a famous example is CeCu$_2$Si$_2$, the
first heavy fermion superconductor\cite{Steglich-CeCu2Si2}.
Therefore it is very interest to study the Ce-based 122-nickel
pnictides like CeNi$_2$As$_2$.

Remarkably, CeNi$_2$As$_2$ crystallizes in either ThCr$_2$Si$_2$
(I4/$mmm$, No.139) or CaBe$_2$Ge$_2$ (P4/$nmm$, No.129) structure,
see Fig.~\ref{Fig.1}(a). The main difference between them comes from the
interchange of the Ni and As positions in one-half of the NiAs
layers in the CaBe$_2$Ge$_2$-type CeNi$_2$As$_2$, which results in
the loss of the inversion symmetry\cite{Jeitschko}. Previous
studies on poly-crystalline samples have revealed that the
ThCr$_2$Si$_2$-type CeNi$_2$As$_2$ shows an antiferromagnetic
(AFM) transition at around 5 K, while the CaBe$_2$Ge$_2$- type
CeNi$_2$As$_2$ is a non-magnetic Kondo lattice
compound\cite{CeNi2X2}. Single crystalline samples are then highly
desirable in order to further distinguish the properties of the
two structures. However, since the occurrence of
ThCr$_2$Si$_2$-type or CaBe$_2$Ge$_2$-type largely depends on the
process of heat treatment\cite{LnNi2As2}, and in many cases, a
mixture of the two will be derived, the properties of
ThCr$_2$Si$_2$-type CeNi$_2$As$_2$ are still not well-understood.

In this article,  we report the measurements on the magnetic,
transport, and thermodynamic properties of the ThCr$_2$Si$_2$-type
CeNi$_2$As$_2$ based on millimeter sized single crystalline
samples. Single crystalline sample of CeNi$_2$As$_2$ was
successfully synthesized by the NaAs flux method. We find that
CeNi$_2$As$_2$ is a highly anisotropic uniaxial antiferromagnet
with the transition temperature $T_N$=4.8 K. The Ce$^{3+}$ moments
are likely to align along $c$-axis. A magnetic field induced
meta-magnetic transition (MMT) was seen below $T_N$. Pronounced
crystalline electric field (CEF) effect was observed. These
magnetic and thermodynamic properties can be well understood by
the CEF calculation, showing that the $j=5/2$ multiplet of
Ce$^{3+}$ splits into three Kramers doublets with the excitation
energies $\Delta_1=$325 K and $\Delta_2=$520 K. In contrast to the
CaBe$_2$Ge$_2$-type CeNi$_2$As$_2$, Kondo effect in the
ThCr$_2$Si$_2$-type CeNi$_2$As$_2$ is not strong, with a
moderately enhanced Sommerfeld coefficient $\gamma_0=$69
mJ/mol/K$^2$ and a relatively low Kondo scale $T_K\sim$ 4 K. On
the other hand, a huge frustration parameter $f=\theta_W/T_N$ is
obtained, and a frustration-distortion picture
 was then proposed. The latter highlights the important role of Ce-$4f$
 electrons in magnetic frustrations. Therefore, the ThCr$_2$Si$_2$-type CeNi$_2$As$_2$
provides a new candidate for the research of frustration-induced
magnetic and structure transitions and calls for more
investigations.

\section{Experimental}

\begin{figure}[htbp]
\includegraphics[width=9cm]{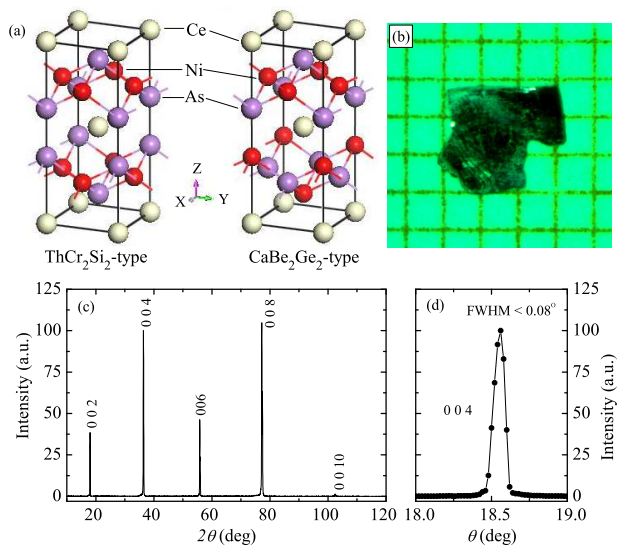}
\caption{(Color online) \label{Fig.1}(a), Crystalline structure of
CeNi$_{2}$As$_{2}$ of ThCr$_2$Si$_2$-type (left) and
CaBe$_2$Ge$_2$-type (right). (b), a photo of a CeNi$_2$As$_2$
single crystal on millimeter grid paper. (c), XRD structure
characterization of CeNi$_2$As$_2$ single crystal. Only (0 0 2$l$)
peaks can be observed. (d), X-ray rocking curve through (0 0 4)
reflection of CeNi$_2$As$_2$ single crystal.}
\end{figure}

High purity single crystal of ThCr$_2$Si$_2$-type
CeNi$_{2}$As$_2$ was grown by NaAs flux method. CeAs, NiAs and
 NaAs were presynthesized as mentioned elsewhere\cite{ZAXu-CeNiAsO1,NaAs}.
 CeAs, NiAs, CeO$_2$ \cite{CeO2-note} and Ni were weighted in the ratio of
1:1:1:1, thoroughly ground in an Argon filled glove box. The mixture
was then put into a Ta tube, together with 15 molar times of NaAs. After
sealing the Ta tube by an Arc-melter, the tube was then sealed into a quartz
tube filled with 0.2 bar Argon gas. The quartz tube was heated up to
1493 K and kept at that temperature for 10 hours, followed by slowly
cooling down to 873 K in 10 days. NaAs flux was dissolved by water
in a fume hood, and shining single crystals of CeNi$_2$As$_2$ with a
typical size of 3 $\times$ 3 $\times$ 0.2 mm$^3$ were picked out from
the remaining dreg (see Fig.~\ref{Fig.1}(b)).

The single crystalline CeNi$_2$As$_2$ samples were checked by X-ray
diffraction (XRD), performed on a PANalytical X-ray diffractometer
(Empyrean Series 2) with Cu-K$_{\alpha 1}$ radiation at room
temperature. Only (0 0 2$l$) peaks can be observed (Fig.~\ref{Fig.1}(c)),
confirming the ThCr$_2$Si$_2$-type crystalline structure. The full
width at half maximum (FWHM) of (0 0 4) peak in the rocking scan is
less than 0.08 \textordmasculine, demonstrating the goodness of
sample quality (Fig.~\ref{Fig.1}(d)). We also performed Rietveld refinement
\cite{Rietan-FP} on the powder XRD data (not shown). The
structural parameters are listed in Tab.~\ref{table1}. The derived occupation
of Ni site is 0.856, close to the result of 0.86 obtained from
energy-dispersive X-ray microanalysis (EDX) measurement. It is also
comparable with previous literature\cite{LnNi2As2}. The derived
$a$($b$) and $c$ are 4.0806(3){\AA} and 9.8843(7){\AA},
respectively. The small ratio $c/a=$2.43, compared to 2.80 for
BaNi$_2$As$_2$ \cite{Sefat-BaNi2As2}, manifests a collapsed ThCr$_2$Si$_2$ structure.

\begin{table}
\tabcolsep 0pt \caption{\label{table1} Lattice parameters of CeNi$_{2}$As$_{2}$
derived from Rietveld refinement based on space group I4/$mmm$.
$a$=$b$=4.0806(3){\AA}, $c$=9.8843(7){\AA},
$\alpha$=$\beta$=$\gamma$=90\textordmasculine. } \vspace*{-12pt}
\begin{center}
\def\temptablewidth{1.0\columnwidth}
{\rule{\temptablewidth}{1pt}}
\begin{tabular*}{\temptablewidth}{@{\extracolsep{\fill}}ccccc}
Atom                          & Occupation  & $x$~   &$y$~   &$z$ \\
\hline
Ce                         &1.000 & 0    & 0     & 0\\
Ni                         &0.856 & 0    & 0.5   & 0.25\\
As                         &1.000 & 0    & 0     & 0.3654(3)\\
\end{tabular*}
{\rule{\temptablewidth}{1pt}}
\end{center}
\end{table}

Quantum Design (QD) magnetic property measurement system (MPMS-5)
and physical property measurement system (PPMS-9) were used in the
magnetization, transport and specific heat measurements. Ohmic
contact was made with Epoxy silver paste (Part A+B), and annealed in
Ar atmosphere at 573 K for 30 min. Resistivities of both in-plane
($\rho_{ab}$, $\textbf{I}\parallel ab$) and out-of-plane
($\rho_{c}$, $\textbf{I}\parallel c$) configurations were measured.
Thermopower was measured by means of steady-state technique and a
pair of differential type E thermocouples was used to measure the
temperature gradient. Specific heat was measured by heat pulse
relaxation method in PPMS-9.

\section{Results and Discussion}

\begin{figure*}[htbp]
\includegraphics[width=15cm]{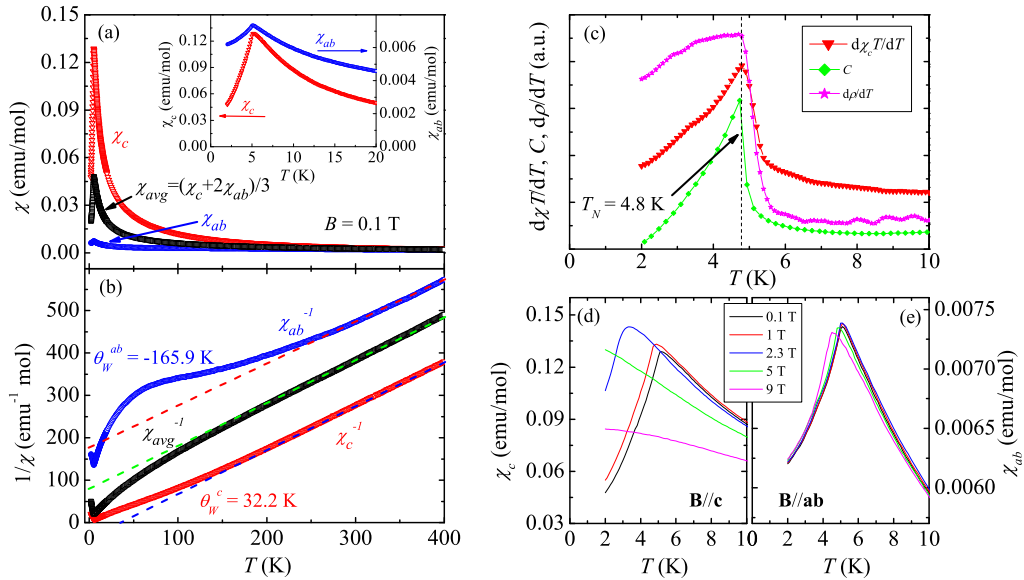}
\caption{(Color online)\label{Fig.2} (a) Temperature dependent magnetic
susceptibility of CeNi$_2$As$_2$ measured at $B=$ 0.1 T shown in
$\textbf{B}\parallel \textbf{c}$ (red) and $\textbf{B}\parallel
\textbf{ab}$ (blue) directions. Polycrystalline averaged
susceptibility (black) was calculated by
$\chi_{avg}=(\chi_c+2\chi_{ab})/3$. (b) shows inverse magnetic
susceptibility. The dashed lines are guides to eyes of the Curie-Weiss fit.
(c) shows the definition of the AFM transition
temperature $T_N$ from $d\chi T/dT$, $C(T)$ and $d\rho/dT$. (d) and
(e) exhibit the evolution of AFM peak in $\chi(T)$ under various
magnetic field, for $\textbf{B}\parallel\textbf{c}$ and
$\textbf{B}\parallel\textbf{ab}$, respectively.}
\end{figure*}

The temperature dependent magnetic susceptibility $\chi(T)=M/H$ and
inverse magnetic susceptibility $1/\chi(T)$ along $\textbf{B}
\parallel \textbf{c}$ and $\textbf{B} \parallel \textbf{ab}$ are displayed in Fig.~\ref{Fig.2}(a) and
(b), respectively. The magnitude of $\chi_c$ is almost the same as that of
$\chi_{ab}$ at 400 K, but is 16 times lager at low temperature,
indicative of increasing magnetic anisotropy. Both $\chi_c(T)$ and
$\chi_{ab}(T)$ obey the Curie-Weiss's law above 300 K. We fit the
temperature dependent susceptibility to the formula
$\chi(T)=\frac{C}{T-\theta_W}$, with $\theta_W$ being the so-called
Weiss temperature. The fit on the polycrystalline averaged
susceptibility, defined as $\chi_{avg}=(\chi_c+2\chi_{ab})/3$, leads
to the effective moment $\mu_{eff}=2.44 \mu_B$. This value is
close to but slightly less than 2.54 $\mu_B$, the effective moment
of a free Ce$^{3+}$, manifesting the trivalent Ce ion and the
non-magnetic nature of the Ni sublattice. The high magnetic
anisotropy is also reflected in the derived Weiss temperature,
$\theta_W^{c}$=32.2 K and $\theta_W^{ab}$=-165.9 K.  An AFM
transition is signified by $\chi_c$, which shows a sharp peak around
5 K and extrapolates to a very small magnitude in the zero
temperature limit. $\chi_{ab}$ also shows a peak at the same
temperature, although the reduction of $\chi_{ab}$ after the formation
of the AFM ordering is much weaker. The characteristic temperature
of the AFM transition, $T_N$=4.8 K, is then determined by the peak
position in the $d\chi T/dT$ curves as shown in Fig.~\ref{Fig.2}(c). We will
find that this value is also consistent with the resistivity
($d\rho/dT$) and specific heat ($C$) measurements. Fig.~\ref{Fig.2}(d) and (e)
show the $\chi_c(T)$ and $\chi_{ab}(T)$ measured under various
magnetic fields. It is interesting to notice that under increasing
field the AFM peak shifts to lower temperatures much more faster for
$\textbf{B}\parallel \textbf{c}$ than $\textbf{B}\parallel
\textbf{ab}$. These observations suggest that the Ce$^{3+}$ moments
aligns along the $c$-axis while within the $ab$-plane the correlation
between the moments is much stronger. The deviation of $\chi(T)$ from
Curie-Weiss law below 300 K is a sign of CEF effect and will be
discussed later on.

\begin{figure}[htbp]
\includegraphics[width=8cm]{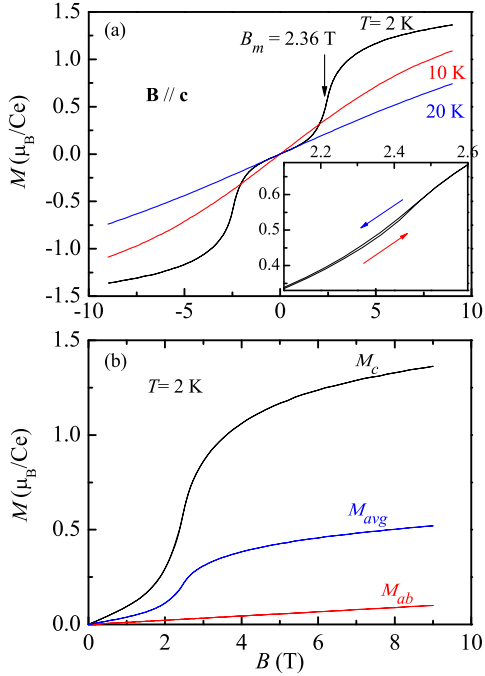}
\caption{(Color online)\label{Fig.3} (a) Field dependence of isothermal
magnetization of CeNi$_2$As$_2$ in $\textbf{B} \parallel
\textbf{c}$, measured at 2 K, 10 K, and 20 K. Inset shows a tiny
hysteresis near the MMT. (b) shows $M(B)$ curves for both
$\textbf{B} \parallel \textbf{c}$ and $\textbf{B} \parallel
\textbf{ab}$ measured at 2 K. The polycrystalline averaged $M(B)$ is
presented in (b).}
\end{figure}

Fig.~\ref{Fig.3} shows isothermal magnetization $M(B)$ curves along
$\textbf{B}\parallel \textbf{c}$ and $\textbf{B} \parallel
\textbf{ab}$ directions. The most fascinating feature for
$\textbf{B} \parallel \textbf{c}$ is that below $T_N$, $M(B)$ shows
linear $B$ dependence when $B<$2 T, but undergoes a substantial
increase at around 2.4 T before a saturation trend. $B_m=$2.36 T is
then defined at the magnetic field where the increasing rate of
$M(B)$ reaches the maximum. A tiny hysteresis in $M(B)$ is observed
near $B_m$, implying a weak first order transition. Compared with
$\textbf{B} \parallel \textbf{c}$, $M(B)$ for $\textbf{B} \parallel \textbf{ab}$
is linear and much smaller, which again provides evidence for the
strong anisotropy in the magnetic correlation among Ce moments. This
field induced MMT may imply the
competition between the in-plane correlation and Zeeman energy.

\begin{figure}[htbp]
\includegraphics[width=8cm]{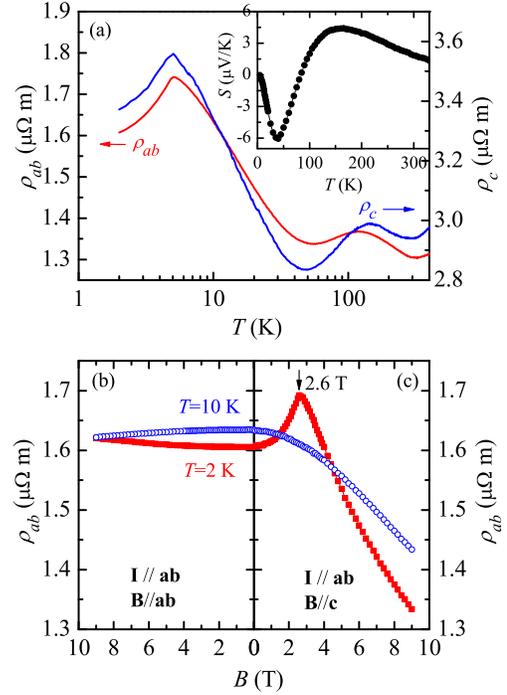}
\caption{(Color online)\label{Fig.4} Transport properties of CeNi$_2$As$_2$. (a)
Mainframe, temperature dependence of resistivity for both
$\rho_{ab}$ ($\textbf{I}\parallel\textbf{ab}$) and $\rho_c$
($\textbf{I}\parallel\textbf{c}$). Inset displays thermopower $S$ as a function of $T$. (b) and (c), $\rho_{ab}$ as a function of magnetic
field $B$, for $\textbf{B}\parallel\textbf{ab}$ and
$\textbf{B}\parallel\textbf{c}$, respectively.}
\end{figure}

We now turn to the resistivity measurement. Both in-plane
($\rho_{ab}$ with $\textbf{I}\parallel\textbf{ab}$) and out-of-plane
($\rho_c$ with $\textbf{I}\parallel\textbf{c}$) resistivities were
measured, and the data are shown in Fig.~\ref{Fig.4}(a). One should notice
that the ratio $\rho_c/\rho_{ab}$ at 400 K is 2.3, much smaller than
that of the regular iron pnictide $A$-122 compounds
\cite{ChenXH-Ba122Ani,WuD-Eu122Ani,WuG-Ca122Ani,Sefat-BaNi2As2,Tomioka-BaNi2P2}
where the ratio is typically of the order of 10-100. It demonstrates
more three-dimensional electronic property in CeNi$_2$As$_2$, and is
consistent with the collapsed crystalline structure. For $T>$300 K,
$\rho_{ab}$ shows weak metallic behavior, while in $T<$300 K region,
$\rho_{ab}$ increases with decreasing $T$, and a broad peak centered
around 110 K is observed. Similar behavior is also observed in
$\rho_{c}$ whereas the broad peak position is relatively higher.
The resistivity of its non-magnetic reference LaNi$_2$As$_2$ was
 also measured (data not shown) on a poly-crystalline sample, and no
 anomaly can be seen at this $T$ region. Such broad peak in resistivity is then reminiscent
 of the CEF effect. Furthermore, a maximum in thermopower (see the inset of Fig.~\ref{Fig.4}
here) observed around 155 K in CeNi$_2$As$_2$ allows us to get a
rough estimate of the first excited Kramers doublet that is
$\sim$310 K above the ground doublet. More detailed and accurate
CEF analysis will be performed on the magnetic susceptibility
fitting, seeing the context below. The anisotropic response to the
CEF in resistivity (see also in Fig.~\ref{Fig.9}(a)) may reflect
the anisotropic hybridization strength of electron scattering to
the CEF. Another prominent feature is observed below 50 K, where
both $\rho_{ab}$ and $\rho_c$ increase rapidly, developing the
sharp peaks near $T_N$. The -$\log T$ behavior of low temperature
resistivity for $T_N<T<$50 K is identified, revealing that
CeNi$_2$As$_2$ belongs to a Kondo system with a weak Kondo scale
$T_K \lesssim T_N$. Fig.~\ref{Fig.4}(b) and (c) show the
isothermal in-plane resistivity $\rho_{ab}$ versus the applied
field perpendicular and parallel to the crystallography $c$-axis,
respectively. In the case of $\textbf{B}
\parallel \textbf{ab}$, $\rho_{ab}$ decreases slightly with $B$ at
$T=10$ K $>T_N$. 
While at $T=2$ K $<T_N$, a positive magnetoresistivity ($MR$,
defined as $MR$=$(\rho(B)-\rho(0))/\rho(0)$) is clearly exhibited.
This behavior is likely associated with the suppression of AFM
ordering under the external field. The $\rho_{ab}(B)$ curves for
$\textbf{B}
\parallel \textbf{c}$ are more intriguing: First, at $T=$10 K,
$\rho_{ab}$ decreases much faster for $\textbf{B}
\parallel \textbf{c}$ than that for $\textbf{B}
\parallel \textbf{ab}$, providing further evidence that the magnetic
easy axis to be $c$-axis.
Second, at $T=$2 K, $\rho_{ab}(B)$ substantially increases to a
maximum near $B=$2.6 T, and then decreases drastically. The turning
point of $\rho_{ab} (B)$ is apparently associated with the MMT
observed in magnetization measurement discussed previously.

\begin{figure}[htbp]
\includegraphics[width=8cm]{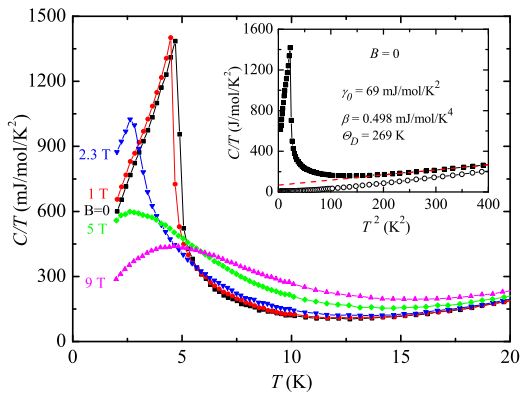}
\caption{(Color online)\label{Fig.5} Specific heat of
CeNi$_2$As$_2$ as a function of temperature, measured at various
magnetic fields $\textbf{B} \parallel \textbf{c}$. Inset displays
$C/T$ versus $T^2$ to estimate the Sommerfeld coefficient
$\gamma_0$. The solid symbols represent CeNi$_2$As$_2$, while open
ones represent LaNi$_2$As$_2$.}
\end{figure}

In Fig.~\ref{Fig.5}, we present the specific heat divided by $T$ as a function
of temperature. A $\lambda$-shape peak is clearly seen at the
transition temperature $T_N$, manifesting a second-order phase
transition. Under the magnetic field, the specific
heat peak is suppressed to lower temperatures for $B<B_m$, indicating a
fingerprint of the reduction of the AFM transition. When $B>B_m$,
the sharp peak evolves with increasing magnetic field into a broad
round peak moving to higher temperature. This Schottky-like peak in
$C/T$ under a field signifies a crossover from AFM ordering to
paramagnetic state via Brillouin-like saturation, and is consistent
with the magnetic properties measurement.
We also plot the $C/T$ vs. $T^2$ in the inset of Fig.~\ref{Fig.5}. The
Sommerfeld coefficien $\gamma_0 \sim$ 69 mJ/mol$\cdot$K$^2$ is then
estimated by linearly extrapolating to the zero temperature. This
moderately enhanced Sommerfeld coefficient manifests the correlation
effect contributed from the Ce-$4f$ electrons. The slope of the
linear extrapolation is $\beta=$0.498 mJ/mol/K$^4$, and this leads
to the Debye temperature $\Theta_D=$269 K. Due to short range order
 or CEF effects mentioned above, this estimate of $\Theta_D$ may has some
 uncertainty. A more reliable estimate of $\Theta_D$ may come from LaNi$_2$As$_2$ (poly-crystal),
  where similar analysis leads to $\Theta_D=$256 K.

In order to identify the Ce-$4f$ electron contribution to the
specific heat, we consider the quantity $C_{mag}=C_{Ce}-C_{La}$ as
shown in Fig.~\ref{Fig.6}(a), where $C_{La}$ is the specific heat of
LaNi$_2$As$_2$. It is expected that the difference
$C_{mag}$ is mainly due to the magnetic contribution because
LaNi$_2$As$_2$ is non-magnetic and isostructural to CeNi$_2$As$_2$.
As expected, a sharp specific heat jump for $C_{mag}$ appears at
$T_N$. The jump at $T_N$ is $\Delta C_{mag}\mid_{T=T_{N}}\sim$ 6
J/mol/K. From this value the Kondo scale could be roughly estimated
as $T_K\sim$ 4 K\cite{Besnus_TK/TN}. The magnetic entropy gain ($S_m$)
was calculated by integrating $C_{mag}/T$ over $T$. We found
that $S_m$ reaches 65\% of $R\ln2$ at $T_N$, and recovers this
value at 19 K. This suggests that the observed anomaly in specific
heat arises from the AFM ordering of Ce$^{3+}$ moment in a two-degenerated
ground state. Since $T_K$ is low, the reduction of
entropy gain at $T_N$ should be mainly attributed to the short-range
ordering or correlation of the Ce$^{3+}$ moments above $T_N$, rather
than the Kondo effect. The short-range ordering is also manifested
by a noticeable broad tail in $C_{mag}$ above $T_N$ extending to
about 20 K (see Fig.~\ref{Fig.6}(a)). A broad peak centered at around 160 K can
also be observed on $C_{mag}(T)$, which should be ascribed to the
Schottky anomaly caused by the thermal population of CEF levels (see also in the
inset of Fig.~\ref{Fig.6}(a), where $S_m$ substantially increases and reaches $R\ln4$ near 150 K).

In Ce-contained compounds, the $D_{4h}$ (I4/$mmm$) point
symmetry requires a CEF Hamiltonian written as:
\begin{equation}
\mathscr{H}_{CEF}=B_2^0 O_2^0+B_4^0 O_4^0+B_4^4 O_4^4,
\label{Eq.1}
\end{equation}
where $B_l^m$ ($l=$2,4, $m=$0,4) are the CEF parameters, while
$O_l^m$ are Steven's operators\cite{Stevens,Hutchings}. In addition,
the Zeeman interaction and exchange interaction should also be taken
into account,
\begin{eqnarray}
\mathscr{H}_{Zee} = -g \mu_B \textbf{J}\cdot\textbf{B},~~~~~~~~~~~~~~~~~~\label{Eq.2}\\
\mathscr{H}_{ex}=-\sum_{<i,j>}
J_{ex}^{\perp}(S_i^xS_j^x+S_i^yS_j^y)+J_{ex}^{\parallel}S_i^zS_j^z,
\label{Eq.3}
\end{eqnarray}
in which $x$, $y$ and $z$ correspond to the crystallographic $a$,
$b$ and $c$, where $g=$6/7 is the Land\'{e} factor of Ce$^{3+}$ ions,
$J_{ex}^{\perp}$ and $J_{ex}^{\parallel}$ are the components of the
nearest-neighbor exchange interaction with Ce$^{3+}$ moment
perpendicular and parallel to the $c$-axis, respectively. Combination
of Eq.~(\ref{Eq.1}), (\ref{Eq.2}) and (\ref{Eq.3}) allows us to get the expressions of inverse
susceptibility\cite{Boutron-CEFint,Canfield-HoNi2B2C}, i.e.,
\begin{eqnarray}
\frac{1}{\chi_{c}}=\frac{1}{C}(T+\frac{j(j+1)}{3}J_{ex}^{\parallel}+\frac{(2j-1)(2j+3)}{5}B_2^0),
\label{Eq.4}\\
\frac{1}{\chi_{ab}}=\frac{1}{C}(T+\frac{j(j+1)}{3}J_{ex}^{\perp}-\frac{(2j-1)(2j+3)}{10}B_2^0).
\label{Eq.5}
\end{eqnarray}
with $j=$5/2 being the total angular momentum for Ce$^{3+}$. The
experimental data of $1/\chi_c$ and $1/\chi_{ab}$ can be well
reproduced by this model as shown in Fig.~\label{Fig.6}(b), with the best fitted
CEF parameters $B_l^m$ as well as the CEF energy levels and
eigenstates being listed in Tab.~\ref{table2}. The ground state has a two-fold
degeneracy and takes the form $\mid \Gamma_7 \rangle= \alpha_1 \mid
\pm \frac{5}{2} \rangle + \alpha_2 \mid \mp \frac{3}{2} \rangle$
with $\alpha_1^2+\alpha_2^2=1$. The calculated exchange interactions
are $J_{ex}^{\parallel}$=9.18 K and $J_{ex}^{\perp}$=-23.1 K,
respectively, manifesting anisotropic magnetic couplings. The broad
peak in $C_{mag}(T)$ can also be well described by Schottky
anomaly formula with energy excitations
$\Delta_1$=325 K and $\Delta_2$=520 K, and the result is shown in
Fig.~\ref{Fig.6}(a). A schematic sketch of the CEF split
is presented in Fig.~\ref{Fig.6}(d). We should point out that, a direct relation\cite{WangYL-CEFPM} of
the paramagnetic Curie temperatures with the CEF parameter B$_2^0$, viz.,
$\theta_W^{ab}-\theta_W^c=\frac{3}{10}B_2^0(2j-1)(2j+3)$ will lead to
$B_2^0=$20.6 K that does not satisfy all the measured physical properties.
We attributed this inconsistency to the neglect of the exchange interactions
which play an important role in determining the ground state of this CEF splitting\cite{Boutron-CEFint}.

\begin{figure}[htbp]
\includegraphics[width=9cm]{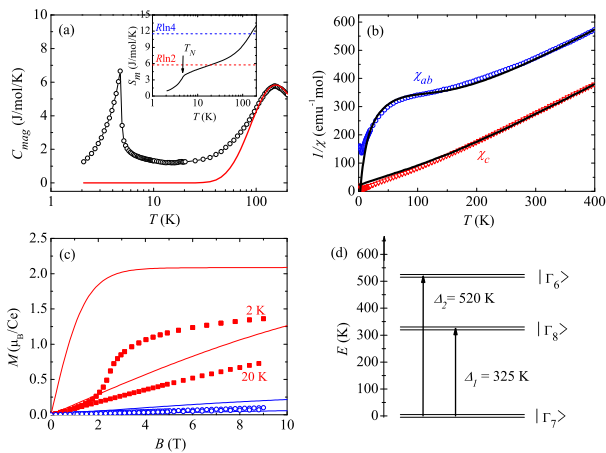}
\caption{(Color online)\label{Fig.6} (a) Magnetic specific heat $C_{mag}$,
derived from $C_{Ce}-C_{La}$. Solid line represents the 3-level
Schottky anomaly fit based on the CEF model, where $\Delta_1=$325 K
and $\Delta_2=$520 K. Inset of (a) shows the integrated entropy as a
function of $T$, the two horizontal dashed lines stand for $R \ln 2$
and $R \ln 4$, respectively. (b) The inverse susceptibility (open symbols)
and calculated curves (solid lines). (c) Experimental (symbols) and
calculated (solid lines) isothermal magnetization at 2 K and 20 K.
Data for $\textbf{B} \parallel \textbf{c}$ are colored with red,
while $\textbf{B} \parallel \textbf{ab}$ are in blue. (d) Schematic
sketch of CEF energy levels for Ce$^{3+}$ ion in CeNi$_2$As$_2$.}
\end{figure}

\begin{table*}
\caption{\label{table2} CEF parameters, energy levels and wave
functions in CeNi$_2$As$_2$ at zero magnetic field.}
\begin{ruledtabular}
\begin{center}
\def\temptablewidth{1.6\columnwidth}
\begin{tabular}{ccccccc}
\multicolumn{6}{l}{CEF parameters } \\
 & $B^{0}_{2}$=-28.2 K, & $B^{0}_{4}$=0.106 K, & $B^{4}_{4}$=1.63 K  \\
 \\ \hline \hline
\multicolumn{6}{l}{Energy levels and Eigenstates}  \\  $E$(K)& $\mid
\frac{5}{2},+\frac{5}{2} \rangle$ & $\mid \frac{5}{2},+\frac{3}{2}
\rangle$ & $\mid \frac{5}{2},+\frac{1}{2} \rangle$ & $\mid
\frac{5}{2},-\frac{1}{2} \rangle$
& $\mid \frac{5}{2},-\frac{3}{2} \rangle$ & $\mid \frac{5}{2},-\frac{5}{2} \rangle$ \\
\hline
0 & -0.9907  & 0      & 0  & 0      & 0.1362  & 0 \\
0 & 0      & -0.1362 & 0       & 0  & 0      & 0.9907 \\
325  & 0.1362 & 0      & 0  & 0      & 0.9907  & 0 \\
325  & 0      & 0.9907 & 0       & 0  & 0      & 0.1362 \\
520     & 0      & 0 & 1       & 0 & 0      & 0 \\
520     & 0  & 0      & 0   & 1      & 0  & 0 \\
\end{tabular}
\end{center}
\end{ruledtabular}
\end{table*}

\begin{figure*}[htbp]
\includegraphics[width=16cm]{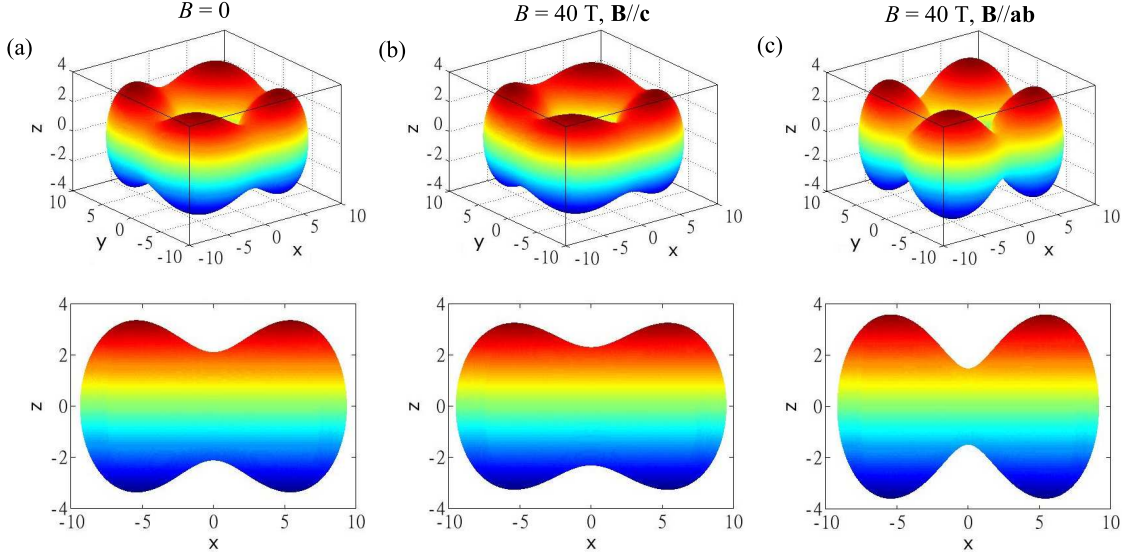}
\caption{(Color online)\label{Fig.7} Iso-surface plot of $4f$-charge density for
an Ce$^{3+}$ ion surrounded by CEF in CeNi$_2$As$_2$. Calculated at
$T=$0.1 K, (a), $B=0$, (b), $B=$40 T,
$\textbf{B}\parallel\textbf{c}$, and (c), $B=$40
T,$\textbf{B}\parallel\textbf{ab}$. The lower diagrams are the
projections to the $xz$-plane.}
\end{figure*}

Above CEF analysis also allows calculating the spatial distribution
of the $4f$-electron charge density\cite{Bauer-CEF}
$\langle\Gamma_7\mid\rho_{4f}(\textbf{r})\mid\Gamma_7\rangle$. In
Fig.~\ref{Fig.7}(a), we display the iso-surface plot (namely the electron
cloud) of the $4f$-electron for an Ce$^{3+}$ ion surrounded by the
CEF in the ThCr$_2$Si$_2$-type CeNi$_2$As$_2$. The calculation was
done at $T=$0.1 K. It is evident that the electron cloud has
deformed severely from the spherical shape, and is highly
accumulated on the four corners which are along the Ce-As bonds,
manifesting the hybridization between Ce-$4f$ and As-$4p$ orbitals.
Another profound feature is that the Ce-$4f$ electron cloud is
highly "squeezed" along the $z$-axis (Please note the dimensions of
$x$-,$y$-, and $z$-axes). It is well known that the topology of charge
density is closely associated with the magnetic anisotropy. Due to
the spin-orbit coupling, the orientation of the magnetic moment is
coupled to the orientation of the $4f$ charge. In the CEF theory,
this anisotropy is mainly governed by the Stevens factor
$\alpha_j$\cite{Hutchings,Bauer-CEF}. In the case of Ce$^{3+}$ ions,
the negative $\alpha_j=$-5.7143$\times$10$^{-2}$ favors the magnetic
easy axis to be parallel along the "squeezed" direction. This result
reinforces the previous statement that the Ce$^{3+}$ moments are
aligned along the $c$-axis. Under a magnetic field $\textbf{B}
\parallel \textbf{c}$, the electron cloud is further "squeezed" along
$z$-axis (see Fig.~\ref{Fig.7}(b)), and thus stabilizes the original magnetic
easy axis. In contrast, an external field $\textbf{B} \parallel
\textbf{ab}$ elongates the electron cloud along the $z$-direction
(see Fig.~\ref{Fig.7}(c)), and consequently the magnetic moments will be
rotated to the $ab$-plane. 

\begin{figure}[htbp]
\includegraphics[width=8cm]{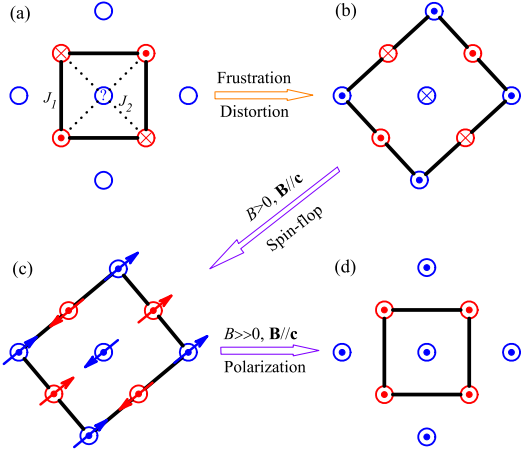}
\caption{(Color online)\label{Fig.8} Schematic diagram of the magnetic structure
of Ce$^{3+}$ sublattice. The adjacent Ce$^{3+}$ layers are denoted
by colors, red and blue, while the orientations of a Ce$^{3+}$
moment, up and down, are signified by "$\bullet$" and "$\times$".
The thick lines characterize the unit cell. (a)Geometry frustration
stems from the $J_1-J_2$ competition in the tetragonal phase, and
tiny structure distortion may take place, which will lead to an
orthorhombic phase as shown in (b). (c) Spin-flop transition happens
when a field $\textbf{B} \parallel \textbf{c}$ is applied, which
enhances the structure distortion. The arrows display the projection
of the Ce$^{3+}$ moment in the $ab$-plane. (d) Geometry frustration
reduces when Ce$^{3+}$ moments are well polarized in the high field
limit.}
\end{figure}

\begin{figure}[htbp]
\includegraphics[width=9cm]{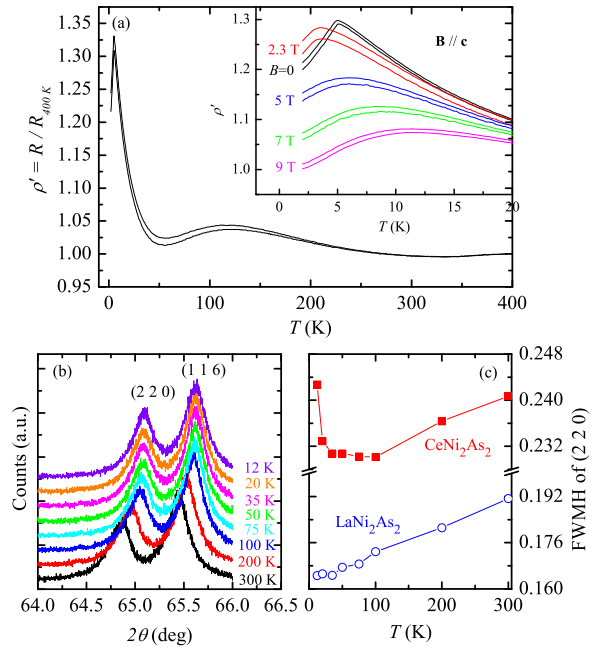}
\caption{(Color online)\label{Fig.9} (a), Anisotropic in-plane resistivity of
CeNi$_2$As$_2$, measured in Van der Pauw's method. Inset displays
this anisotropy under various field. (b) LTXRD patterns of
CeNi$_2$As$_2$ powders. The two observed peaks are indexed as
(2~2~0) and (1~1~6). (c) Temperature dependence of FWHM of (2~2~0)
peak. For comparison, the result of LaNi$_2$As$_2$ is also shown. }
\end{figure}

However, we notice that the CEF model does not perfectly reproduce
the isothermal magnetization measured at low temperatures, see in
Fig.~\ref{Fig.6}(c). For example, in the case of $\textbf{B} \parallel
\textbf{c}$, one expects a saturated magnetic moment
$M_c$=6/7$\times$5/2=2.14 $\mu_B$/Ce, while the experimental value
is about 1.36 $\mu_B$/Ce for $T=2$ K and $B=$9 T. Meanwhile, we also
notice that the amplitude of Weiss temperatures are much larger than the AFM
transition temperature. Especially for the in-plane Weiss
temperature $\theta_W^{ab}$, we obtain a huge ratio
$f_{ab}=\theta_W^{ab}/T_N$=34.6. Such a large value of $f_{ab}$
reminds us of the magnetic frustration neglected in the previous CEF
analysis. A schematic diagram of this magnetic frustration is shown
in Fig.~\ref{Fig.8}. Taking into account only the nearest neighbor exchange
interactions for both intralayer and interlayer, the magnetic
coupling between Ce$^{3+}$ moments are denoted by $J_1$ and $J_2$
respectively. From the CEF calculation we have a negative intralayer
coupling which is dominating in magnitude. It means that the
Ce$^{3+}$ moments should be antiferromagnetically ordered within the
$ab$-plane, as shown in Fig.~\ref{Fig.8}(a). In this situation, the magnetic
frustration stems from the $J_1$-$J_2$ competition for the moments
in the two adjacent layers as within the extended unit cell
\cite{DaiJH-frustration}. We propose that a structural distortion
from the high temperature tetragonal to low temperature orthorhombic
phases may possibly take place to release this magnetic
frustration, as shown in Fig.~\ref{Fig.8}(b). The unit cell is then doubled and
the lattice constants $a$ and $b$ are no longer equivalent. This
frustration-induced distortion scenario is reminiscent to a similar
scenario in the iron pnictides, where the structural distortion is
possibly due to the $J_1$-$J_2$ magnetic frustration caused by the
$3d$-electron moments\cite{WangHu2008,Xu2008,Dai-PNAS}. What we need
to emphasize here is that the proposed frustration-induced
distortion in the present case is caused by the Ce-$4f$ electrons.
With this consideration, the CEF Hamiltonian in the low temperature
orthorhombic phase can be written as
\begin{equation}
\mathscr{H}_{CEF}=B_2^0 O_2^0+B_2^2 O_2^2+B_4^0 O_4^0+B_4^2
O_4^2+B_4^4 O_4^4.
\label{Eq.6}
\end{equation}
Consequently, the ground state will change into a mixed state like
$\mid \Gamma \rangle_0$=$\alpha_1 \mid \pm \frac{5}{2}\rangle +
\alpha_2 \mid \mp \frac{3}{2}\rangle +\alpha_3 \mid \pm
\frac{1}{2}\rangle$ with $\alpha_1^2+\alpha_2^2+\alpha_3^2$=1, and
therefore a saturated magnetic moment smaller than 2.14
$\mu_B$/Ce will be expected under a moderate magnetic field. This
situation is similar to the case in CeNiGe$_3$ reported by Mun
et al\cite{Canfield-CeNiGe3}. However, it is hard to get the $B_l^m$'s
for this new phase, since the information is very limited.

To obtain more evidences for the structural distortion, we sought for
the possible resistivity anisotropy in the $ab$-plane by performing
the $\rho_{a'}$-$\rho_{b'}$ measurement using the Van der Pauw's
method\cite{VanderPauw} (Please note that here $\textbf{a'}$ and $\textbf{b'}$
stand for two perpendicular directions in the $ab$-plane, not
necessarily the crystallographic $\textbf{a}$ and $\textbf{b}$).
The measurement was carried out on a piece
of square plate-like single crystal, and $R_{a'}$ and $R_{b'}$ were
measured via switching the direction of electrical current. To
compare their temperature dependences, the normalized resistivity
$\rho' = R /R_{400 K}$ is used, and the result is displayed in
Fig.~\ref{Fig.9}(a). Above 200 K, $\rho'_{a'} (T)$ and $\rho'_{b'} (T)$ overlap
well, while below 200 K, a discrepancy between them is observed. It
should be pointed out that this result was reproduced for many times
on different batches of single crystals. The discrepancy becomes
more evident with decreasing temperature, exhibiting the increasing
anisotropy in the $ab$-plane. 
It is worth emphasizing that this discrepancy between $\rho'_{a'}$ and
$\rho'_{b'}$ can be enlarged under a moderate magnetic field
$\textbf{B}
\parallel \textbf{c}$ and reaches a maximum near $B_m$ before it
starts to decrease with further increased $B$ (see the inset of
Fig.~\ref{Fig.9}(a)). Combined with the MMT observed in isothermal
magnetization displayed in Fig.~\ref{Fig.3}, we argue that such field
dependent $\rho'_{a'}$-$\rho'_{b'}$ may be related to the field induced
spin-flop transition. As is elucidated in Fig.~\ref{Fig.8}(c), under a
moderate magnetic field $\textbf{B} \parallel \textbf{c}$, the
balance between the Zeeman energy and the AFM intralayer coupling
requires that the Ce$^{3+}$ magnetic moments gradually lie down to
the $ab$-plane. This will then further enhance the structural distortion as
well as the resistivity anisotropy in the $ab$-plane. When the
external field is large enough, all the magnetic moments tend to be
polarized, and thus the frustration decays with the increasing
field.

The frustration-distortion scenario can be further tested by the low
temperature X-ray diffraction (LTXRD) experiment on the
CeNi$_2$As$_2$ powder samples. We were focused on the angular range
64 $\textordmasculine \leq 2\theta \leq$ 66 $\textordmasculine$,
where only the (2~2~0) and (1~1~6) peaks can be observed. The data
were collected at different temperatures down to 12 K, the lowest
temperature of our equipment. All the collected LTXRD patterns
are displayed in Fig.~\ref{Fig.9}(b). We find that both (2~2~0)
 and (1~1~6) peaks shift to the right-hand-side when cooling down,
suggesting a shrinkage of the crystalline lattice, although for
$T<$100 K, this shrinkage becomes very weak. We fit all these LTXRD
patterns to a combination of two Gaussian functions, through which
the FWHM of (2~2~0) peak is derived as shown in Fig.~\ref{Fig.9}(c). The
initial reduction of the FWHM with decreasing temperature should be
attributed to the slowing down of crystalline lattice oscillation.
To our interest, an upturn of the FWHM is clearly seen when $T<$ 100
K, and especially for $T<$20 K, the FWHM increases rapidly with
decreasing temperature, although we are not able to see the split of
(2~2~0) peak directly. Such steep increase of the FWHM signals a
tiny structural distortion or a precursor to that happened below
$T_N$. In order to confirm the relevance of this behavior to the
$4f$-electron magnetism, we also measured the FWHM of the
LaNi$_2$As$_2$ compound, see in Fig.~\ref{Fig.9}(c). We find that the FWHM of
LaNi$_2$As$_2$ drops monotonically with $T$ down to the lowest
temperature, in striking contrast to CeNi$_2$As$_2$. The significant
distinction betweens the two cases again reinforces the
frustration-distortion possibility driven by the $4f$-electrons.

We should remark that the structural distortion induced by the
magnetic frustration of $4f$-electrons is in general small or even
tiny. One reason is that the energy scale for magnetic couplings
is smaller than that in the iron pnictides. Another reason is that
the next-nearest-neighbor coupling $J_3$ between the Ce-moments
(within the same layer of the Ce-sublattice as denoted in
Ref.\cite{DaiJH-frustration}), which is neglected in the previous
discussions, may also play some role. Though being small, $J_3$
usually competes with both $J_1$ and $J_2$ and may lead to
incommensurate magnetic fluctuations\cite{DaiJH-frustration}.
Since a precise magnetic structure in CeNi$_{2}$As$_{2}$ is not
easy to be determined by static magnetization and transport
measurements, more investigations such as neutron scattering
experiment are required to settle this issue.

Finally, we note that CeNi$_2$P$_2$, the counterpart compound to
CeNi$_2$As$_2$, behaves as a typical Kondo lattice
metal\cite{CeNi2X2}. This fact demonstrates that the chemical
"pressure", induced by replacing As with smaller isovalent P
\cite{Dai-PNAS,XuZA-CeFeAsPO}, promisingly acts as an effective
controlling parameter to tune the competition between the RKKY
interaction and Kondo coupling\cite{Doniach}. For the rare earth iron pnictides,
such competition is much involved and complicated, however, mainly
due to the emergent magnetic order of the iron
$3d$-electrons\cite{DaiJH-frustration}. Owing to the
absence of magnetism in the Ni sub-lattice, the Ce-Ni based
compounds have shown great advantage in studying the Ce-$4f$
electron
correlation\cite{Canfield-CeNiGe3,ZAXu-CeNiAsO1,ZAXu-CeNiAsO2,Pikul-CeNi2Ge2}.
This also accounts to the fact that the CaBe$_2$Ge$_2$-type
CeNi$_2$As$_2$ is a non-magnetic Kondo lattice, because the
Kondo coupling is largely enhanced by one of the inverted NiAs
layer. Compared with all these cases, the ThCr$_2$Si$_2$-type
CeNi$_2$As$_2$ has a relatively small Kondo coupling but a
moderately strong magnetic frustration. The role played by strong
magnetic frustration on the quantum phase transition, in Kondo
lattice in particular, remains an interesting issue\cite{Si-Solid,
Coleman-Low,SeijiSi-Low}. Therefore, the ThCr$_2$Si$_2$-type
CeNi$_2$As$_2$ may provide a new material for the research
of quantum phase transitions mediated by the $4f$-electron magnetic
frustration.

\section{Conclusion}

To conclude, we performed a systematic investigation on the magnetic
properties and the CEF effect in the ThCr$_2$Si$_2$-type
CeNi$_2$As$_2$ single crystals. We find that this CeNi$_2$As$_2$
compound is a highly anisotropic uniaxial antiferromagnet with
$T_N$=4.8 K. The Kondo effect is estimated to be not strong in this
system, while the magnetic frustration of the Ce-$4f$ moments plays
an important role. Pronounced CEF effect is observed in magnetic,
transport and thermodynamic measurements. Detailed calculations
based on the CEF theory allows capturing the electronic and
magnetic properties of CeNi$_2$As$_2$.  A possible
frustration-induced structural distortion due to the Ce-$4f$
electrons is suggested, which is in agreement with the in plane resistivity
anisotropy and low-temperature XRD measurements. While this issue is
reminiscent of the frustration-induced distortion emergent in the
iron pnictide superconductors due to the $d$-electron correlation,
the origin of the structural distortion and its relationship with
the magnetic frustration in the $4f$-electron systems still need to
be clarified in the future.

\section*{Acknowledgments}

Y. Luo would like to show gratitude to Hui Xing and Chao Cao for
helpful discussions. This work was supported by the National Basic
 Research Program of China (Grant Nos. 2011CBA00103, 2012CB927404 and 2010CB923003),
 the National Science Foundation of China (Grant Nos. 11190023 and 11174247),
 and the Fundamental Research Funds
for the Central Universities of China. J. Dai was also supported
by the Natural Science Foundation of Zhejiang Province (Grant No.
Z6110033).


\begin{thebibliography}{10}
\bibitem{CeO2-note}
The actual role of CeO$_2$ here is still unknown. We have tried this
synthesis without CeO$_2$, but the product turns out to be of
CaBe$_2$Ge$_2$ type CeNi$_2$As$_2$. Besides, the same synthesis on
LaNi$_2$As$_2$ single crystal also failed.
\end{thebibliography}
\end{document}